\documentclass{article} 
\usepackage{iclr2027_conference,times}

\usepackage{amsmath,amsfonts,bm}

\def\eqref#1{equation~\ref{#1}}

\def\1{\bm{1}}

\DeclareMathAlphabet{\mathsfit}{\encodingdefault}{\sfdefault}{m}{sl}
\SetMathAlphabet{\mathsfit}{bold}{\encodingdefault}{\sfdefault}{bx}{n}

\usepackage{amssymb}
\usepackage{hyperref}
\usepackage{url}
\usepackage{booktabs}
\usepackage{subcaption}
\usepackage{makecell}
\usepackage[table]{xcolor}
\usepackage{tabularx}
\usepackage{amsmath,booktabs,graphicx,array,makecell,multirow,colortbl,xcolor}
\makeatletter
\patchcmd{\iclrfinalcopy}{}{}{}{}
\makeatother

\title{ProxyEraseAgent: Blind Watermark Removal in the Wild}

\author{
Jun Yao$^{1}$,
Chao Wang$^{1}$,
Yupeng Qiu$^{2}$,
Zehua Ma$^{1}$,
Weiming Zhang$^{1}$,
Bin Liu$^{1 \dagger}$,
Han Fang$^{1\dagger}$
\\
$^{1}$University of Science and Technology of China\\
$^{2}$ National University of Singapore \\
\texttt{\{flowice@ustc.edu.cn, fanghan@ustc.edu.cn\}}
}

\begin{document}
\iclrfinalcopy

\maketitle

\providecommand{\methodname}{\textsc{ProxyEraseAgent}}

\begin{abstract}
Invisible image watermark removal has received growing attention. Despite substantial progress, existing attacks still face a tension between practicality and attack specificity. Attacks that exploit detector outputs, decoder responses, or paired watermarked and clean images can be tailored to the watermark decision boundary, but they require information that is rarely available in realistic dissemination scenarios. In contrast, attacks based on compression, geometric distortion, or image reconstruction are easy to deploy from a single watermarked image, but they are largely open-loop: they apply generic transformations without knowing whether the current image is moving toward watermark failure. Thus, the key challenge in single-image blind watermark removal is not merely how to transform the image, but how to obtain a useful direction for removal without access to the hidden decoder. To bridge this gap, we propose \textbf{ProxyEraseAgent}, an agent-driven framework that recovers attack specificity through proxy decoder responses. Publicly available watermarking schemes provide a natural external knowledge base of candidate decoders, where some are informative for a given unknown watermarked image. Our insight is that a decoder producing a stronger calibrated response to the query image is more likely to share a nearby decoding boundary with the hidden target mechanism. ProxyEraseAgent ranks knowledge-base decoders by calibrated response strength and uses the top ones as proxy boundary estimators. Their responses then guide a progressive search over heterogeneous removal operations, including geometric distortion, JPEG compression, image reconstruction, and gradient-based perturbation, under perceptual-quality constraints. Experiments across 11 image watermarking systems show that ProxyEraseAgent achieves an attack success rate of \textbf{94.8\%}, demonstrating the effectiveness of response-guided proxy retrieval and feedback-driven sequential planning for blind watermark removal.

\end{abstract}

\section{Introduction}
Invisible image watermarking has become an important technique for content provenance, copyright protection, and synthetic-image identification~\citep{fernandez2023stable,an2024waves,zhu2018hidden,jia2021mbrs}. As watermarked images are increasingly disseminated through open platforms, watermark removal attacks have become an important means of evaluating whether watermarking systems remain robust under realistic misuse and redistribution~\citep{an2024waves}. In this work, we study a common and practical attack scenario: an attacker observes a disseminated watermarked image and aims to make its hidden watermark unrecoverable while preserving the visible content. Unlike settings where the attacker can query a detector, access a decoder, or exploit knowledge of the watermarking algorithm, this scenario provides only the image itself. This raises a key question: \textit{how can a removal attack remain specific to the hidden watermark when the attacker has access only to the disseminated image?}

The difficulty of this setting does not lie in the absence of possible image transformations. Compression, filtering, geometric distortion, reconstruction, and perturbation can all be readily applied to a watermarked image~\citep{an2024waves,zhao2024invisible,saberi2024robustness,kassis2025unmarker}. The real challenge is that the hidden watermark is not directly accessible: from image appearance alone, the attacker cannot tell whether a candidate operation has changed the embedded message, weakened the decoding signal, or moved the image closer to watermark failure. As a result, a visually acceptable transformation may leave the watermark intact, while a stronger transformation may unnecessarily degrade image quality without providing additional disruption. Therefore, the key challenge in single-image blind removal is not merely how to transform the image, but how to build a prior-free yet closed-loop removal process that can adapt its operations according to attack-time signals.

Existing watermark removal attacks can be broadly understood from this perspective. Feedback-aware attack rely on target-specific prior knowledge, such as detector outputs, decoder responses, gradients, paired watermarked-clean images, or target-specific surrogate models, to estimate whether a transformation is disrupting the watermark~\citep{jiang2023evading,hu2025transfer}. Such information provides a direct estimate of removal progress, but is rarely available when the attacker only observes a disseminated watermarked image. In contrast, transformation-driven attack avoid such assumptions and applies generic image operations, such as compression, geometric distortion, denoising, regeneration, or detector-free removal~\citep{an2024waves,zhao2024invisible,saberi2024robustness,kassis2025unmarker,shamshad2026raven}. These attacks do not require target-specific prior knowledge, but they are largely open-loop: they transform the image without estimating whether the hidden watermark has been meaningfully weakened. This reveals a fundamental dilemma: existing attacks either estimate removal progress using unavailable target-specific prior knowledge, or remain prior-free but open-loop.

The key to resolving this tension is to recover attack-specific guidance without relying on the hidden target decoder. Ideally, one would identify the exact watermarking algorithm and use its decoder to guide removal, but this is unrealistic when only a single watermarked image is available. We take a different perspective: blind removal does not require identifying the exact watermarking algorithm. Instead, if we can infer from the query image which external decoder behaves similarly to the hidden target decoder, that decoder may provide a useful removal direction. This makes publicly available watermarking schemes valuable. Although most candidates will not match the target mechanism exactly, their responses to the query image can reveal which ones are more likely to approximate the hidden boundary.

Based on this idea, we propose ProxyEraseAgent, an agent-driven framework for single-image blind watermark removal. ProxyEraseAgent turns a feedback-free removal problem into a proxy-guided decision process through two coupled stages. First, it performs response-guided proxy decoder retrieval. Our key insight is that a decoder producing a stronger calibrated response to the query image is more likely to provide an informative proxy for the hidden decoding boundary. Accordingly, ProxyEraseAgent evaluates candidate decoders from the public watermark knowledge base, calibrates their response strengths using self-encoded references, and selects the top-ranked decoders as proxy boundary estimators. Second, it uses the responses of these proxy decoders to guide removal path planning. Instead of committing to a fixed transformation, ProxyEraseAgent progressively explores a heterogeneous attack space, including geometric distortion, JPEG compression, image reconstruction, and gradient-based perturbation, and selects the attack path that most effectively suppresses proxy responses while satisfying perceptual-quality constraints. In this way, decoder responses play a dual role: they identify informative proxy decoders and provide the feedback signal needed to plan an attack-specific removal path without querying the hidden target decoder.


\begin{figure}[t]
    \centering
\includegraphics[width=1.0\textwidth]{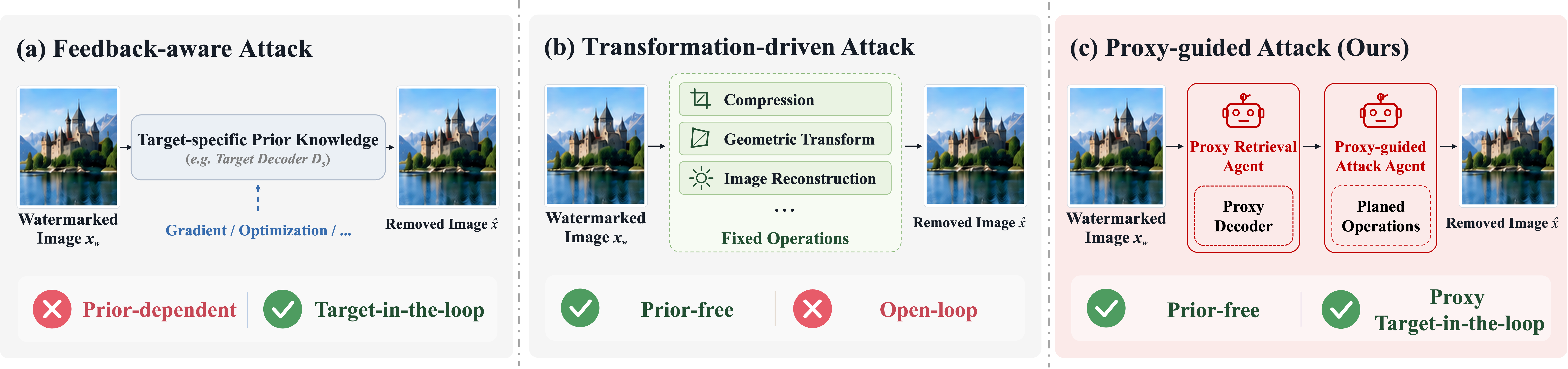}
    \caption{
    Comparison of different paradigms for invisible image watermark removal.
    }
    \label{fig:method_comparison}
\end{figure}

Experiments across 11 image watermarking systems show that ProxyEraseAgent achieves an attack success rate of \textbf{94.8\%}, demonstrating that response-guided proxy retrieval and sequential attack planning can provide effective attack-specific guidance without querying the hidden target decoder.

Our main contributions are summarized as follows:
\begin{itemize}

\item We identify a gap in blind watermark removal: existing attacks are either prior-dependent or prior-free but open-loop, motivating a proxy-in-the-loop formulation for single-image blind removal.

\item We propose \textbf{ProxyEraseAgent}, which retrieves informative proxy decoders by calibrated responses and uses their outputs to guide sequential removal planning under perceptual-quality constraints.

\item We evaluate ProxyEraseAgent on 11 watermarking systems, achieving a \textbf{94.8\%} attack success rate, with ablations confirming the contributions of proxy decoder retrieval and sequential attack planning.

\end{itemize}

\section{Related Work}
\label{sec:related_work}

\paragraph{Invisible image watermarking.}
Deep image watermarking has developed from end-to-end encoder--decoder models toward systems tailored to different robustness requirements and deployment scenarios. HiDDeN~\citep{zhu2018hidden} established a general framework for embedding and decoding hidden messages in images, while MBRS~\citep{jia2021mbrs}, CIN~\citep{ma2022towards}, FIN~\citep{fang2023flow}, and PIMoG~\citep{fang2022pimog} improve robustness against compression, noise, black-box distortions, and screen-capture effects, respectively. Other systems extend watermarking to latent representations, provenance tracking, arbitrary resolutions, and image--video applications, including RoSteALS~\citep{bui2023rosteals}, InvisMark~\citep{xu2025invismark}, TrustMark~\citep{bui2023trustmark}, VideoSeal~\citep{fernandez2024video}, and ChunkySeal~\citep{petrov2025we}. This diversity in embedding mechanisms, payload capacities, and robustness objectives makes watermark behavior highly heterogeneous, so an attack that is effective for one system may transfer poorly to another.

\paragraph{Watermark removal.}
Existing watermark removal attacks can be broadly divided into two paradigms. The first applies predefined image transformations, including geometric transformations and JPEG compression~\cite{an2024waves}, denoising or purification~\cite{saberi2024robustness}, adversarial perturbations~\cite{jiang2023evading}, and generative reconstruction~\cite{zhao2024invisible}. Benchmarking frameworks such as WAVES~\cite{an2024waves} integrate multiple attack families to systematically evaluate watermark robustness, while diffusion- and reconstruction-based methods can effectively disrupt pixel-domain watermarks while preserving much of the visible image content~\cite{zhao2024invisible,saberi2024robustness}. However, such attacks are typically executed with predefined operators, strengths, or orders, providing little guidance on which operation is most effective for a particular image or target system. As a result, they are difficult to adapt to heterogeneous watermarking mechanisms on a per-image basis. 
A second line of work introduces surrogate models to provide attack guidance when the target decoder is unavailable. WEvade optimizes adversarial perturbations on a surrogate detector and transfers them to target watermark detectors~\cite{jiang2023evading}. More recent work further improves transferability by independently training multiple surrogate watermarking models with different architectures or configurations and jointly optimizing attacks across them~\cite{hu2025transfer}. Although surrogate-based approaches reduce the need for direct access to the target model, they still rely on additional surrogate-model training, auxiliary data, or a predefined surrogate set.  In contrast, our method does not train dedicated surrogate models from scratch. It treats publicly available watermark decoders as an external knowledge base, retrieves informative proxy decoders according to their calibrated responses to the query image, and uses the resulting proxy feedback to guide adaptive sequential search over heterogeneous attack operations.


\section{Method}

\begin{figure}[t]
    \centering
    \includegraphics[width=\textwidth]{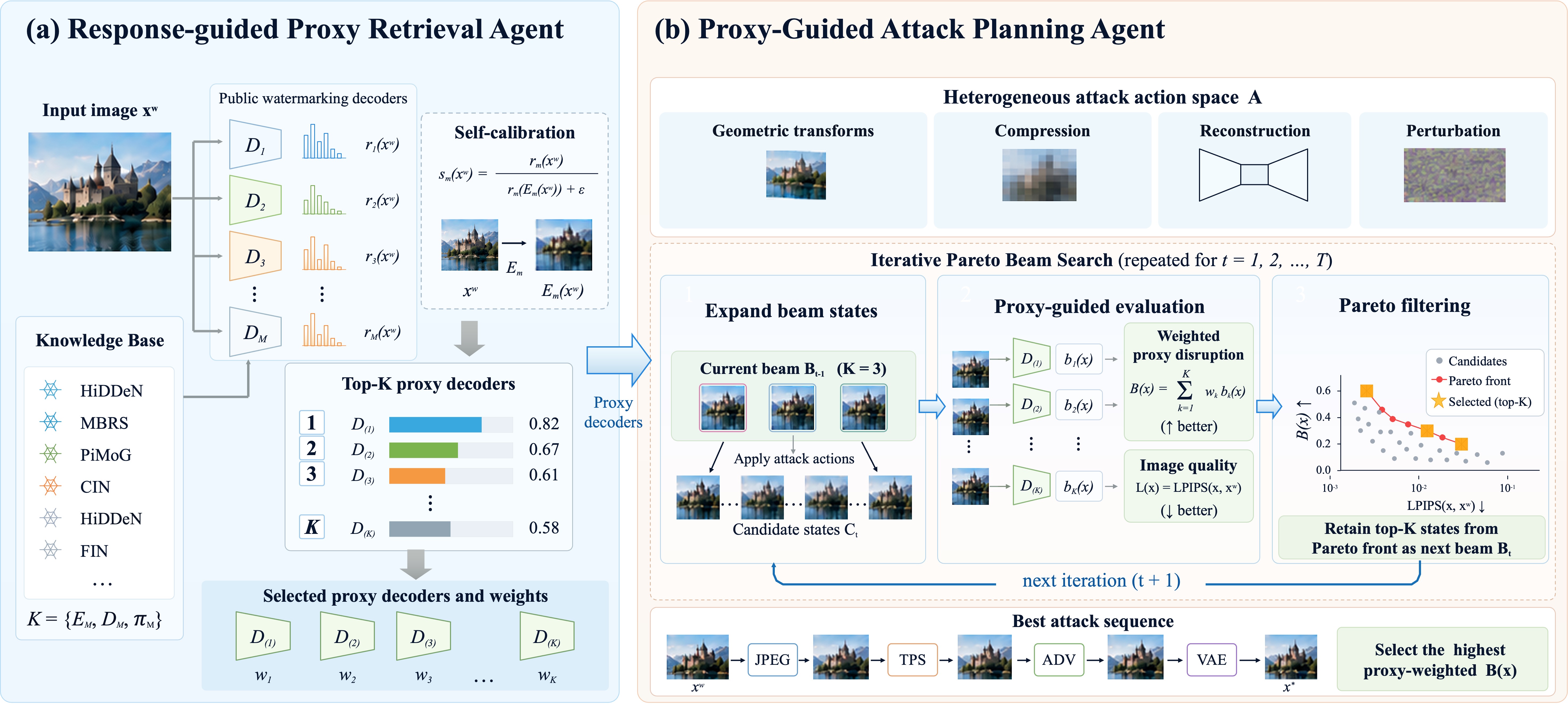}
    \caption{
   Overview of ProxyEraseAgent. The Response-guided Proxy Retrieval Agent retrieves informative decoders from knowledge base, while the Proxy-Guided Attack Planning Agent uses their feedback to construct an adaptive watermark-removal path. The target decoder is never accessed during proxy retrieval or attack search and is used only for final evaluation.
    }
    \label{fig:method_pipeline}
\end{figure}

\subsection{System Overview}
We propose ProxyEraseAgent, a system for single-image blind watermark removal. As illustrated in Figure~\ref{fig:method_pipeline}, ProxyEraseAgent consists of three main components: a heterogeneous decoder knowledge base, a Response-guided Proxy Retrieval Agent for identifying informative proxy decoders, and a Proxy-Guided Attack Planning Agent for constructing adaptive watermark-removal paths.
Given a watermarked image $x^{w}$, the Response-guided Proxy Retrieval Agent evaluates candidate decoders in the knowledge base and ranks them according to their calibrated responses to the input. The top-ranked decoders are selected as proxy decoders and provide feedback for subsequent attack planning. Based on these proxy signals, the Proxy-Guided Attack Planning Agent performs sequential search over a heterogeneous attack space and adaptively composes attack operations to maximize watermark disruption while controlling perceptual distortion. 

\subsection{Knowledge Base}
To obtain external watermark-related feedback without accessing the target decoder, we construct a heterogeneous knowledge base from publicly available image-watermarking systems. We denote the knowledge base as
\[
\mathcal{K}
=
\left\{
(E_m,D_m,\pi_m)
\right\}_{m=1}^{M},
\]

where $E_m$ and $D_m$ denote the encoder and decoder of the m-th watermarking system, respectively, and $\pi_m$ contains the system-specific metadata required for inference, including input preprocessing, image resolution, message length, and decoding configuration.

The decoder $D_m$ provides watermark-related responses for proxy retrieval and subsequent attack guidance, while the corresponding encoder $E_m$ is used to construct self-watermarked references for response calibration. By incorporating watermarking systems with diverse embedding and decoding mechanisms, the knowledge base provides heterogeneous candidate signals for an unknown target watermark. The specific watermarking systems and their configurations are described in appendix.


\subsection{Response-guided proxy retrieval}
\label{3.3}
Not every decoder in the knowledge base provides useful feedback for a given query image. Therefore, we need to retrieve a small subset of decoders whose responses to $x^{w}$ are most informative.
For an input image $x$, let $p_{m,j}(x)$ denote the probability that the $j$-th watermark bit is predicted as 1 by decoder $D_m$, where $j\in\{1,\ldots,L_m\}$ and $L_m$ is the message length of decoder $D_m$. We quantify the decoding uncertainty using the mean binary entropy \[H_m(x)
=
-\frac{1}{L_m}
\sum_{j=1}^{L_m}
\left[
p_{m,j}(x)\log_2 p_{m,j}(x)
+
\bigl(1-p_{m,j}(x)\bigr)
\log_2\bigl(1-p_{m,j}(x)\bigr)
\right].\]
Here, $H_m(x)$ measures the average uncertainty of decoder $D_m$ over all decoded bits. Since binary entropy is normalized to $[0,1]$ under base-2 logarithms, a smaller $H_m(x)$ indicates more confident bit predictions.
We define the decoder response strength as
$
R_m(x)=1-H_m(x),
$
where a larger $R_m(x)$ indicates more confident bit predictions. Since raw response strengths may have different scales across heterogeneous decoders, we calibrate each decoder using a self-watermarked reference. Specifically, the corresponding encoder $E_m$ embeds a valid watermark into $x^{w}$, producing $E_m(x^{w})$. The calibrated retrieval score is defined as
\[S_m(x^{w})
=
\frac{R_m(x^{w})}
{R_m(E_m(x^{w}))+\varepsilon},
\]

where the numerator measures the response of decoder \(D_m\) to the unknown watermarked image, the denominator measures its response to a reference image watermarked by its own encoder, and \(\varepsilon\) is a small constant for numerical stability. This self-calibration normalizes decoder-specific response scales and makes the responses of heterogeneous decoders more comparable. We rank all eligible candidate decoders according to their calibrated retrieval scores and retain the top-$K$ decoders as proxy decoders. The corresponding retrieval scores are normalized and used as importance weights during attack search.

\subsection{Proxy-Guided Attack Planning Agent}
\label{3.4}
Given the proxy decoders retrieved in the previous stage, the Proxy-Guided Attack Planning Agent constructs an image-specific watermark-removal path by sequentially searching over a heterogeneous attack space. Rather than applying a single predefined transformation, the agent evaluates candidate attack states using proxy-decoder feedback and perceptual quality, and progressively retains promising attack sequences.

\subsubsection{Attack Action Space}
We construct a heterogeneous attack space consisting of four complementary families of watermark-removal operations: geometric distortion, compression distortion, regeneration, and adversarial perturbation. 
\[
\{
A_{\mathrm{geometric}},
A_{\mathrm{compression}},
A_{\mathrm{regeneration}},
A_{\mathrm{adversarial}}
\}.
\]
These attack families capture different mechanisms for disrupting embedded watermark signals and exhibit complementary effectiveness across watermarking systems. Each family contains multiple attack strengths, allowing the Proxy-Guided Attack Planning Agent to jointly determine the operation type and attack strength at each search step. The specific implementations and parameter configurations used in our experiments are provided in Section~\ref{sec:4.1}.







\subsubsection{Proxy-Guided Sequential Attack Search}
Different watermarking mechanisms exhibit substantially different vulnerabilities, making it difficult for a single globally fixed transformation to perform well across all systems. ProxyEraseAgent therefore searches for an image-specific sequence of attack operations.
Starting from $ x_0=x^{w}$,an attack sequence
$
(a_1,\ldots,a_t)
$
recursively generates
$
a_t\circ a_{t-1}\circ\cdots\circ a_1(x^{w}).
$
Each candidate state is evaluated along two dimensions: proxy watermark disruption and perceptual distortion. Because the original watermark message is unavailable, we use the decoding result of each proxy on the original watermarked image as its reference. Let $\hat{\mathbf{y}}_m(x)$ denote the binary message decoded by proxy decoder $D_m$ from image $x$. We define the proxy bit error rate (BER) to measure the degree of disruption introduced by a candidate state $x_t$:
\[b_m(x_t)
=
\operatorname{BER}
\left(
\hat{\mathbf{y}}_m(x_t),
\hat{\mathbf{y}}_m(x^{w})
\right).\]

A larger $b_m(x_t)$ indicates that the attack causes a greater change in the message decoded by proxy decoder $D_m$. At the same time, perceptual distortion relative to the original watermarked image is measured by Learned Perceptual Image Patch Similarity 
$\operatorname{LPIPS}(x_t,x^{w})$.

\subsubsection{Pareto Beam Search}
\label{sec:pareto_search}

We use Pareto beam search to construct the attack sequence. Starting from the original watermarked image, each retained state is expanded at every search step using all admissible attack operations and their parameter settings. The resulting candidates are evaluated according to the proxy disruption score $B(x)$ and perceptual distortion $L(x)$.

We first discard candidates whose perceptual distortion exceeds the predefined search budget. We then perform Pareto filtering on the remaining candidates. Specifically, a candidate $x_i$ is dominated by another candidate $x_j$ if
$B(x_j) \geq B(x_i),
L(x_j) \leq L(x_i)$,
with at least one strict inequality. Dominated candidates are removed, and only non-dominated candidates are retained.

To balance attack effectiveness and image fidelity while controlling the search cost, we maintain a Pareto beam of width three at each search depth, retaining representative candidates with different effectiveness--fidelity trade-offs. These candidates are further expanded in the next search step. The search proceeds for at most $T$ steps. After completing the search, we select the candidate with the highest proxy disruption score among all explored states satisfying the perceptual-quality constraint:
\[
x^{*}
=
\arg\max_{x:\,L(x)\leq\tau_{\mathrm{select}}}
B(x),
\]
where $\tau_{\mathrm{select}}=0.05$ in our experiments. Ties are resolved in favor of lower perceptual distortion and then shorter attack sequences.

\section{Experiments}
\label{sec:experiments}

\subsection{Experimental setup}
\label{sec:4.1}

\paragraph{Datasets.}
We evaluate ProxyEraseAgent on 100 randomly selected real-world images from the MS-COCO 2017 dataset~\citep{lin2014microsoft}. 
To evaluate its robustness across heterogeneous watermarking mechanisms, each image is independently watermarked by 11 representative image-watermarking systems, including CIN, ChunkySeal, FIN, HiDDeN, InvisMark, LightweightMark, MBRS, PIMoG, RoSteALS, VideoSeal, and TrustMark.
We use their publicly released models and official inference configurations. Detailed preprocessing, image resolutions, payload lengths, and decoding settings are provided in Appendix~\ref{A1}.


\paragraph{Attack baselines.} 
We compare ProxyEraseAgent with representative watermark-removal attacks from four categories. 
Non-geometric baselines include BM3D, brightness and contrast adjustment, Gaussian blur, Gaussian noise, and JPEG compression, while geometric baselines include cropping, rotation, and scaling. 
For regeneration-based attacks, we evaluate Regen-VAE and Regen-Diffusion~\citep{zhao2024invisible}. 
We further compare with dedicated watermark-removal methods, including AdvEmbB-RN18, AdvEmbB-CLIP~\citep{an2024waves}, and UnMarker~\citep{kassis2025unmarker}. 
For UnMarker, we use the HL configuration following the perceptual-quality constraint in our evaluation. 
Detailed configurations and parameter ranges are provided in Appendix~\ref{attack_config}.


\begin{table*}[t]
    \centering
    \caption{Constrained attack success rate (cASR, \%) for globally fixed attacks and \methodname{}. A success requires target-decoder BER $\geq 0.1$ and AlexNet LPIPS $\leq 0.05$. }
    \label{tab:fixed_attack_main}

    \begingroup

    \definecolor{tablerow}{RGB}{246,246,255}
    \definecolor{tableours}{RGB}{232,233,250}
    \definecolor{tablerule}{RGB}{145,145,160}

    \arrayrulecolor{tablerule}
    \renewcommand{\arraystretch}{1.16}
    \setlength{\tabcolsep}{2pt}

    \footnotesize

    \newcommand{\attackcolwidth}{1.75cm}
    \newcommand{\modelcolwidth}{0.89 cm}
    \newcommand{\overallcolwidth}{1.05cm}

    \newcommand{\colhead}[1]{%
        \parbox[c][1.10cm][c]{\linewidth}{%
            \centering\footnotesize\bfseries
            #1%
        }%
    }

    \begin{tabular}{
        @{}
        >{\raggedright\arraybackslash}p{\attackcolwidth}
        !{\color{tablerule}\vrule width 0.4pt}
        *{11}{>{\centering\arraybackslash}p{\modelcolwidth}}
        !{\color{tablerule}\vrule width 0.4pt}
        >{\centering\arraybackslash}p{\overallcolwidth}
        @{}
    }

        \toprule

        \multicolumn{1}{@{}c!{\color{tablerule}\vrule width 0.4pt}}{%
            \parbox[c][1.10cm][c]{\attackcolwidth}{%
                \centering\footnotesize\bfseries
                Attack type%
            }%
        }
        & \colhead{CIN}
        & \colhead{Chunky\\[-2pt]Seal}
        & \colhead{FIN}
        & \colhead{HiD\\[-2pt]DeN}
        & \colhead{Invis\\[-2pt]Mark}
        & \colhead{Light\\[-2pt]weight}
        & \colhead{MBRS}
        & \colhead{PIMoG}
        & \colhead{RoSte\\[-2pt]ALS}
        & \colhead{Video\\[-2pt]Seal}
        & \colhead{Trust\\[-2pt]Mark}
        & \colhead{Overall\\[-2pt]cASR $\uparrow$}
        \\

        \midrule

        \multicolumn{13}{@{}l@{}}{%
            \textbf{\textit{Non-geometric distortions}}
        }
        \\

        \rowcolor{tablerow}
        Brightness
        & 0.0 & 0.0 & 0.0 & 0.0 & 0.0 & 0.0
        & 0.0 & 0.0 & 0.0 & 0.0 & 0.0 & 0.0
        \\

        \rowcolor{tablerow}
        Contrast
        & 0.0 & 0.0 & 0.0 & 0.0 & 0.0 & 0.0
        & 0.0 & 0.0 & 0.0 & 0.0 & 0.0 & 0.0
        \\

        Gauss. Blur
        & 0.0 & 0.0 & 0.0 & 0.0 & 0.0 & 0.0
        & 0.0 & 0.0 & 0.0 & 0.0 & 0.0 & 0.0
        \\

        \rowcolor{tablerow}
        Gauss. Noise
        & 0.0 & 91.0 & 0.0 & 8.0 & 97.0 & 0.0
        & 88.0 & 0.0 & 0.0 & 88.0 & 68.0 & 40.0
        \\
       BM3D
        & 74.0 & 77.0 & 65.0 & 69.0 & 77.0 & 75.0
        & 79.0 & 17.0 & 3.0 & 72.0 & 70.0 & 61.6
        \\
        JPEG
        & 22.0 & 95.0 & 93.0 & 92.0 & 95.0 & 3.0
        & 95.0 & 61.0 & 43.0 & 95.0 & 96.0 & 71.8
        \\

        \specialrule{1.0pt}{1.5pt}{1.5pt}

        \multicolumn{13}{@{}l@{}}{%
            \textbf{\textit{Geometric distortions}}
        }
        \\

        \rowcolor{tablerow}
        Crop
        & 14.0 & 23.0 & 16.0 & 12.0 & 10.0 & 16.0
        & 28.0 & 0.0 & 17.0 & 2.0 & 0.0 & 12.5
        \\
        \rowcolor{tablerow}
        Scaling
        & 0.0 & 73.0 & 5.0 & 8.0 & 59.0 & 0.0
        & 64.0 & 0.0 & 0.0 & 5.0 & 0.0 & 19.5
        \\
        Rotation
        & 19.0 & 63.0 & 76.0 & 62.0 & 13.0 & 76.0
        & 83.0 & 0.0 & 0.0 & 1.0 & 0.0 & 35.7
        \\
    \specialrule{1.0pt}{1.5pt}{1.5pt}
  \multicolumn{13}{@{}l@{}}{%
            \textbf{\textit{Dedicated removal attacks}}
        }
        \\
        \rowcolor{tablerow}
         Unmarker
          & 0.0 & 98.0 & 0.0 & 36.0 & 97.0 & 0.0
        & 96.0 & 0.0 & 3.0 & 97.0 & 42.0 & 42.64
        \\
        AdvB-CLIP
        & 0.0 & 100.0 & 0.0 & 3.0 
        & 82.0 & 0.0 & 85.0 & 0.0 & 0.0 & 100.0 & 98.0&42.5
        \\
        AdvB-RN18
        & 0.0 & 100.0 & 0.0 & 3.0 & 77.0 
        & 0.0 & 91.0 & 0.0 & 2.0 & 100.0 & 98.0& 42.8
        \\
        \rowcolor{tablerow}
        \specialrule{1.0pt}{1.5pt}{1.5pt}
        \multicolumn{13}{@{}l@{}}{%
            \textbf{\textit{Regeneration-based attacks}}
        }
        \\

        \rowcolor{tablerow}
        Regen-Diff
        & 14.0 & 23.0 & 14.0 & 11.0 & 14.0 & 14.0
        & 29.0 & 13.0 & 14.0 & 26.0 & 15.0 & 17.0
        \\
        Regen-VAE
        & \textbf{99.0} & 96.0 & 92.0 & 98.0 & 99.0
        & \textbf{99.0} & 96.0 & 2.0 & 3.0 & 96.0 & 95.0 & 79.5
        \\

        \specialrule{1.0pt}{1.5pt}{1.5pt}

        \rowcolor{tableours}
        \parbox[c][0.85cm][c]{\attackcolwidth}{%
    \raggedright\bfseries
    ProxyErase\\[-1pt]
    Agent%
}
        & 91.0
        & \textbf{100.0}
        & \textbf{97.0}
        & \textbf{100.0}
        & \textbf{100.0}
        & 94.0
        & \textbf{100.0}
        & \textbf{79.0}
        & \textbf{82.0}
        & \textbf{100.0}
        & \textbf{100.0}
        & \textbf{94.8}
        \\

        \bottomrule

    \end{tabular}

    \arrayrulecolor{black}
    \endgroup
\end{table*}

\paragraph{Evaluation metrics.}
We evaluate watermark removal effectiveness using bit error rate (BER) and image quality using LPIPS with an AlexNet backbone. The primary metric is constrained attack success rate (cASR), which considers an attack successful when the target-decoder BER is at least 0.1 while LPIPS is no greater than 0.05:
\[
\mathrm{cASR}=\frac{1}{N}\sum_{i=1}^{N}
\mathbb{I}\left[
\mathrm{BER}_i\geq0.1
\land
\mathrm{LPIPS}_i\leq0.05
\right].
\]



\paragraph{Knowledge-base configuration.}
We construct the decoder knowledge base using publicly available image-watermarking systems, including HiDDeN, MBRS, PIMoG, CIN, FIN, LightweightMark,
RoSteALS, TrustMark, InvisMark, VideoSeal, and ChunkySeal. Detailed preprocessing, image resolutions, message lengths, and decoding settings are provided in Appendix~\ref{Knowledge-base configuration}.

\paragraph{Attack-space configuration.}
Our attack space consists of four operation families: geometric distortion, compression, regeneration, and adversarial perturbation. Specifically, we consider TPS warping for geometric distortion, JPEG compression for compression, VAE reconstruction for regeneration, and a lightweight adversarial optimization attack. Detail are provided in Appendix~\ref{attack-space}.

\subsection{Main Comparison with Watermark-Removal Baselines}
\label{sec:main_results}
We first compare ProxyEraseAgent with representative watermark-removal baselines, including generic transformations, reconstruction-based attacks, and dedicated watermark-specific attacks. As shown in Table~\ref{tab:fixed_attack_main}, ProxyEraseAgent achieves an overall cASR of 94.8\% across 11 watermarking systems, substantially outperforming all baseline methods.
The strongest fixed attack, Regen-VAE, achieves an overall cASR of 79.5\%, while ProxyEraseAgent improves the success rate by 15.3\% under the same perceptual-quality constraint.

The performance of fixed attacks varies substantially across different watermarking systems, indicating that watermark vulnerabilities are highly mechanism-dependent.
For example, Regen-VAE achieves strong performance on several watermarking systems, but its cASR drops to only 3\% on RoSteALS and 2\% on PIMoG.
Similarly, dedicated watermark-specific attacks achieve high effectiveness on some systems but exhibit limited transferability to others.
In contrast, ProxyEraseAgent achieves the best performance on 9 out of 11 watermarking systems, demonstrating more consistent effectiveness across heterogeneous watermarking mechanisms.

\begin{figure}[t]
    \centering
\includegraphics[width=1.0\textwidth]{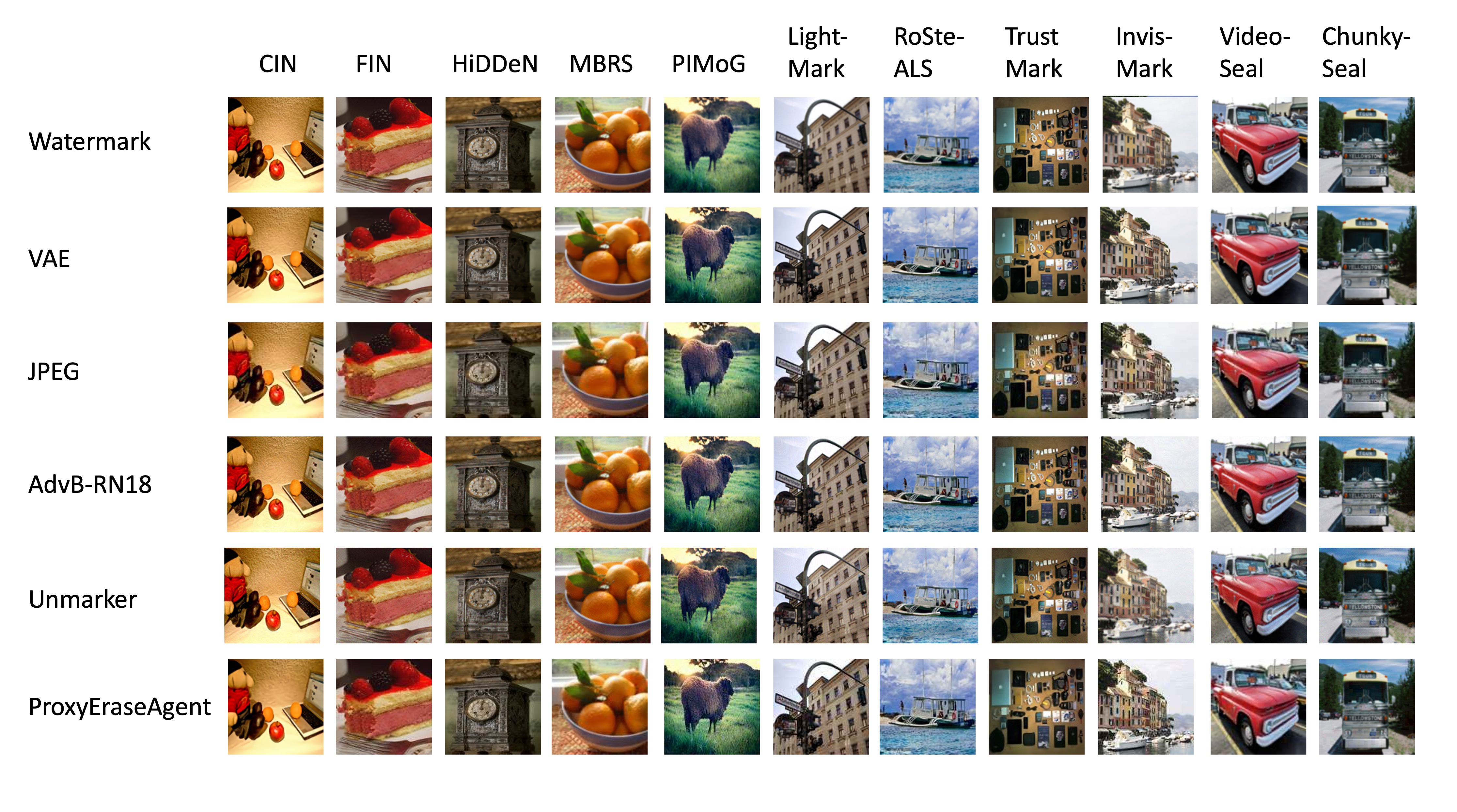}
    \caption{Examples of watermark image attacks, including JPEG compression, VAE-based reconstruction, AdvB-RN18 adversarial attack, Unmarker adversarial attack, and ProxyEraseAgent. 
    }
    \label{fig:baseline_picture}
\end{figure}

These results indicate that a single predefined attack operation is insufficient for blind watermark removal under diverse watermarking schemes.
By retrieving informative proxy decoders and adaptively composing heterogeneous attack operations according to proxy feedback, ProxyEraseAgent can construct image-specific removal paths and achieve stronger attack performance while maintaining perceptual quality. Some attacking examples are shown in Figure ~\ref{fig:baseline_picture}.

\begin{table}[!htbp]
    \centering
    \caption{Ablation studies of ProxyEraseAgent.}
        \begin{subtable}[t]{0.48\textwidth}
        \centering
        \small
        \caption{Ablation
study of proxy decoder selection.}
    \label{tab:candidate_ablation}
        \setlength{\tabcolsep}{3.5pt}
        \begin{tabular}{llcrr}
            \toprule
            Selection  & $K$ & Target & cASR (\%) \\
            \midrule
            Random  & 3 & Excluded & 89.00 \\
            Entropy  & 1 & Excluded & 90.36 \\
            Entropy  & 2 & Excluded & 94.55 \\
            Entropy  & 3 & Excluded & 94.82 \\
            Entropy  & 5 & Excluded & \textbf{95.36} \\
            \midrule
            Entropy  & 3 & Included & 99.73 \\
            \bottomrule
        \end{tabular}
    \end{subtable}
    \begin{subtable}[t]{0.48\textwidth}
        \raggedright
        \small
        \caption{Ablation of maximum search depth.}
        \label{tab:depth_ablation}
        \setlength{\tabcolsep}{2.0pt}
        \begin{tabular}{cccccc}
            \toprule
            Depth & cASR (\%) $\uparrow$ & BER $\uparrow$ &
            LPIPS $\downarrow$ & PSNR $\uparrow$ & Length \\
            \midrule
            1 & 83.91 & 0.3253 & 0.0360 & 30.62 & 1.00 \\
            2 & 92.72 & 0.3649 & 0.0406 & 29.90 & 1.91 \\
            3 & 94.45 & 0.3706 & 0.0418 & 29.59 & 2.51 \\
            4 & \textbf{94.82} & \textbf{0.3745} &
            0.0421 & 29.46 & 2.96 \\
            \bottomrule
        \end{tabular}
    \end{subtable}
    \hfill
\end{table}
\subsection{Effect of candidate-decoder selection}
\label{sec:candidate_ablation}
To evaluate the effectiveness of response-guided proxy decoder retrieval, we analyze the impact of decoder selection strategy and the number of retrieved proxy decoders.
As shown in Table~\ref{tab:candidate_ablation}, under the blind setting where the target decoder is excluded from the knowledge base, randomly selecting three proxy decoders achieves a cASR of 89.00\%.
In contrast, selecting proxy decoders according to calibrated entropy responses achieves a cASR of 94.82\% with $K=3$, demonstrating that response-guided retrieval can identify more informative proxy decoders from a heterogeneous watermark knowledge base.

We further investigate how the number of retrieved proxy decoders affects attack performance.
Using only the highest-ranked proxy decoder ($K=1$) achieves a cASR of 90.36\%, while increasing the number of proxy decoders to $K=3$ improves the cASR to 94.82\%.
This improvement indicates that aggregating feedback from multiple proxy decoders provides more robust guidance and reduces the dependence on any individual decoder.
Increasing $K$ to 5 further improves the cASR to 95.36\%, suggesting that additional proxy signals can still provide marginal benefits, although the improvement gradually saturates. To evaluate the upper bound of ProxyEraseAgent's performance under the condition of complete knowledge base coverage, we assume the knowledge base is sufficiently large to include the target decoder in the pool of candidate decoders. Under this setting, the cASR increases to 99.73\% with $K=3$.

\subsection{Effect of sequential search depth}
\label{sec:depth_ablation}
To evaluate the benefit of sequential attack composition, we vary the maximum search depth of ProxyEraseAgent and analyze the resulting attack performance.
A depth of one corresponds to selecting a single attack operation, while larger depths allow the planner to compose multiple complementary operations into an adaptive removal path.
As shown in Table~\ref{tab:depth_ablation}, restricting the search depth to one achieves a cASR of 83.91\%.
Increasing the depth to two substantially improves cASR to 92.72\%, while further increasing the depth to three and four achieves 94.45\% and 94.82\%, respectively.

The results demonstrate that the effectiveness of ProxyEraseAgent does not come solely from selecting an appropriate single attack operation.
For many watermark systems, a single transformation is insufficient to achieve effective removal under the perceptual-quality constraint, whereas combining multiple complementary operations provides additional disruption capability.
This observation is reflected by the average selected sequence length, which increases from 1.00 at depth one to 2.96 at depth four, indicating that the planner actively utilizes multi-step attack paths when beneficial.

Meanwhile, increasing the search depth introduces only a limited additional perceptual cost.
The average LPIPS increases from 0.0360 to 0.0421 as depth grows from one to four, while remaining within the predefined quality constraint.
Considering the trade-off between attack effectiveness and search complexity, we adopt a maximum search depth of four in the default configuration.

\subsection{Robustness across BER Thresholds}
\begin{figure}[t]
    \centering
\includegraphics[width=1.0\textwidth]{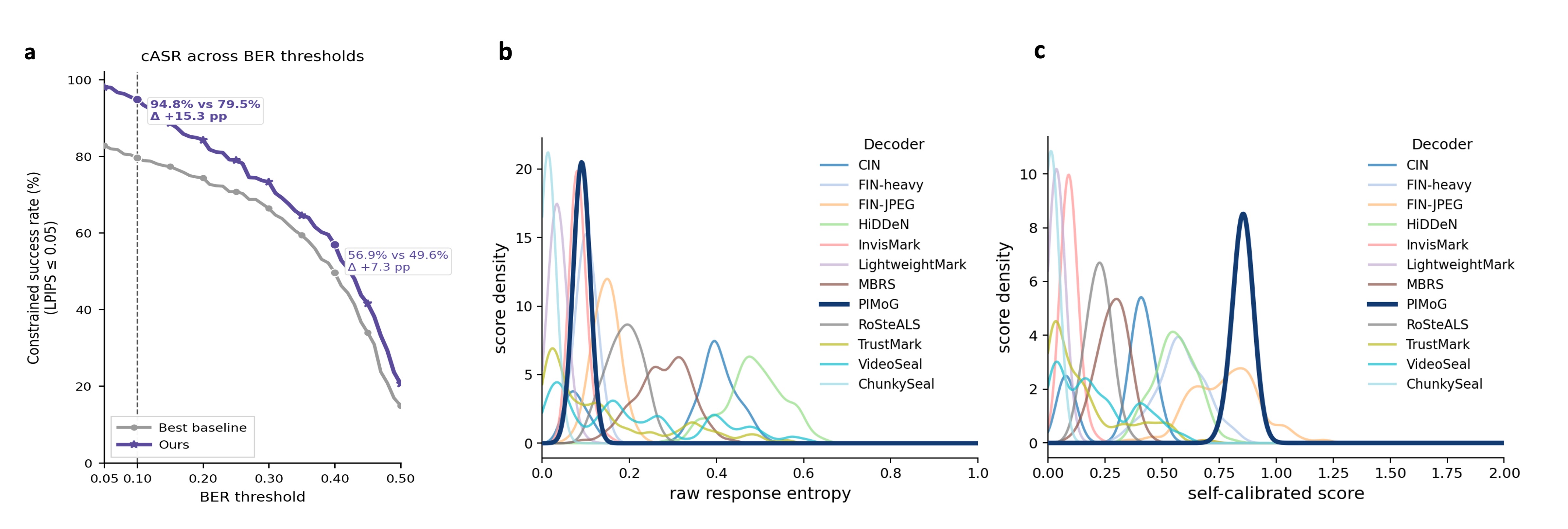}
    \caption{(a) Constrained attack success rate under different BER thresholds. (b) Distribution of raw decoder response entropy. (c) Distribution of self-calibrated retrieval scores.
    }
    \label{fig:ber_threshold}
\end{figure}

To examine whether the effectiveness of ProxyEraseAgent depends on the specific BER threshold used for evaluation, we recompute cASR under different watermark-disruption criteria while keeping the perceptual constraint fixed at $\mathrm{LPIPS}\leq0.05$. As shown in Figure~\ref{fig:ber_threshold}, increasing the BER threshold gradually reduces the attack success rates of all methods, since stricter criteria require stronger watermark disruption.
Nevertheless, ProxyEraseAgent consistently maintains a clear performance advantage over baseline methods across different BER thresholds.
This demonstrates that the improvement of ProxyEraseAgent is not limited to the default threshold of $\mathrm{BER}\geq0.1$, but remains robust under stricter definitions of successful watermark removal.




\subsection{Analysis of Self-Calibrated Retrieval Scores}
\label{sec:calibration_ablation}

To analyze the effect of self-calibration in proxy decoder retrieval, we compare the distributions of raw response entropy and the proposed self-calibrated scores, as shown in Figure~\ref{fig:ber_threshold}.
Without calibration, different watermark decoders exhibit substantially different response ranges, making their raw confidence scores difficult to compare directly. 
After self-calibration, decoder responses are normalized according to their own self-watermarked references, producing a more comparable score space for proxy retrieval.
For example, PIMoG exhibits a relatively narrow raw entropy distribution, which limits its discriminability during decoder ranking, while the calibrated score provides a more distinguishable response distribution.
These results demonstrate that self-calibration reduces decoder-specific confidence bias and enables more reliable proxy decoder selection across heterogeneous watermarking systems.

\section{Conclusion}

We propose \textbf{ProxyEraseAgent}, an agent-driven framework for single-image blind watermark removal without access to the target decoder. ProxyEraseAgent retrieves informative proxy decoders from publicly available watermarking models using calibrated responses, and leverages their feedback to adaptively construct effective attack paths. Experiments across 11 watermarking systems show that ProxyEraseAgent achieves an overall cASR of 94.8\%, outperforming the strongest fixed attack by 15.3 \%. More importantly, our results demonstrate that the key challenge of blind watermark removal can be addressed by recovering informative feedback from external watermarking models and using it to guide adaptive, image-specific attack planning. This provides a new perspective for evaluating watermark robustness in realistic settings where the underlying watermarking mechanism is unknown and inaccessible.




\bibliography{iclr2027_conference}
\bibliographystyle{iclr2027_conference}

\section{Appendix}

\subsection{Watermark Generation Settings.}
\label{A1}
All input images were converted to RGB format and resized to \(128\times128\) before watermark embedding. CIN used the standard configuration with a 30-bit payload and Identity noise. FIN used the HEAVY configuration with a 64-bit payload. HiDDeN used the dropout-0.55--0.6 configuration with a 30-bit payload and the default dropout preprocessing. InvisMark used the standard configuration with a 100-bit payload and internally resized the input to \(256\times256\). MBRS used the EC\_42 configuration with a 256-bit payload, a \(256\times256\) model input, and Identity noise. PIMoG used the default configuration with a fixed 30-bit payload and ScreenShooting distortion. RoSteALS used the VQ4-mir configuration, with the string ``secrets'' encoded as the payload, and processed the input internally at \(256\times256\) resolution. TrustMark used model variant P with BCH-5 encoding, a 61-bit binary payload, and a watermark strength of 1.0. LightweightMark used the DO configuration with a 64-bit payload and a \(128\times128\) input. VideoSeal and ChunkySeal used 256-bit and 1024-bit payloads, respectively. For methods operating at an internal resolution higher than \(128\times128\), the watermarked outputs were resized back to \(128\times128\) after embedding to enable consistent downstream comparison.

\subsection{Attack-space configuration}
\label{attack-space}
\paragraph{Geometric distortion.}
We use thin-plate-spline (TPS) warping with deformation scales
$s\in
\{
2.5\times10^{-4},
5\times10^{-4},
10^{-3},
2\times10^{-3}
\}$.
\paragraph{Compression distortion.}
We use JPEG compression with quality factors
$q\in\{90,80,70,60,50,40,30\}$.

\paragraph{Regeneration}
We use the Cheng2020-anchor model for VAE-based image reconstruction, with quality levels
$ q\in\{1,2,3,4,5,6\}$.

\paragraph{Adversarial perturbation.}
Our framework additionally supports a lightweight low-frequency adversarial operation inspired by UnMarker. The attack combines a mean-pooling loss, $\mathcal{L}_{\mathrm{MPL}}$, which encourages coherent changes in regional image statistics, with a local filter regularization loss, $\mathcal{L}_{\mathrm{FRL}}$, which suppresses abrupt local artifacts:

\[
\mathcal{L}_{\mathrm{low}}
=
-\mathcal{L}_{\mathrm{MPL}}
+
0.25\,\mathcal{L}_{\mathrm{FRL}}.
\]

The optimization jointly updates a pixel-level residual and learnable edge-aware filters while applying appearance regularization to preserve image quality. We use
$
n \in \{50, 100, 200\}
$
optimization iterations.

\subsection{Knowledge-base configuration.}
\label{Knowledge-base configuration}
We construct the decoder knowledge base using publicly available image-watermarking systems, including HiDDeN~\citep{zhu2018hidden}, MBRS~\citep{jia2021mbrs}, PIMoG~\citep{fang2022pimog}, CIN~\citep{ma2022towards}, FIN~\citep{fang2023flow}, Lightweight-Mark~\citep{qiu2025lightweight},
RoSteALS~\citep{bui2023rosteals}, TrustMark~\citep{bui2023trustmark}, InvisMark~\citep{xu2025invismark}, VideoSeal~\citep{fernandez2024video}, and ChunkySeal~\citep{petrov2025we}. These systems cover diverse embedding mechanisms, architectures, payload configurations, and robustness objectives, providing a heterogeneous pool of candidate proxy decoders. We use the publicly released models with their corresponding inference configurations. 

\subsection{Attack baselines onfigurations}
\label{attack_config}
Regarding the parameterization of conventional attacks, we follow the settings used in InvisMark~\citep{zhao2024invisible}. Specifically, the baseline attack set includes brightness adjustments with factors \([2,4,6,8,12]\), contrast adjustments with factors \([2,3,4,5,6,7]\), JPEG compression with quality levels \([10,20,30,40,50,60]\), additive Gaussian noise with zero mean and standard deviations \([5,10,15,20,25,30]\), and Gaussian blur with radii \([2,4,6,8,10,12]\). In addition, we evaluate VAE-based compression at quality levels \([6,5,4,3,2,1]\) and diffusion-based perturbations with noise steps \([0,5,10,20,40,60,80]\).For regeneration-based attacks, we evaluate Regen-VAE and Regen-Diffusion proposed by ~\citep{zhao2024invisible}.  Finally, we compare against dedicated watermark-removal attacks, including the black-box adversarial embedding attacks  AdvEmbB-RN18 and AdvEmbB-CLIP ~\citep{an2024waves}, as well as UnMarker. For UnMarker ~\citep{kassis2025unmarker}, we use its HL configuration rather than the default CHL configuration, since CHL introduces substantially larger image distortion and often exceeds the perceptual-quality constraint used in our evaluation.

\subsection{Attack-search configuration.}
ProxyEraseAgent retrieves the top $K=3$ proxy decoders according to the calibrated response scores described in Section~\ref{3.3}. The Proxy-Guided Attack Planning Agent performs Pareto beam search with a beam width of three and a maximum search depth of four. At each search step, candidate operations and their strength configurations are drawn from the attack space described above. The final attack result is selected under the perceptual-quality constraint of $\mathrm{LPIPS}\leq0.05$. Throughout the search process, the target decoder remains inaccessible and is used only for final evaluation.

\begin{figure}[t]
    \centering
\includegraphics[width=1.0\textwidth]{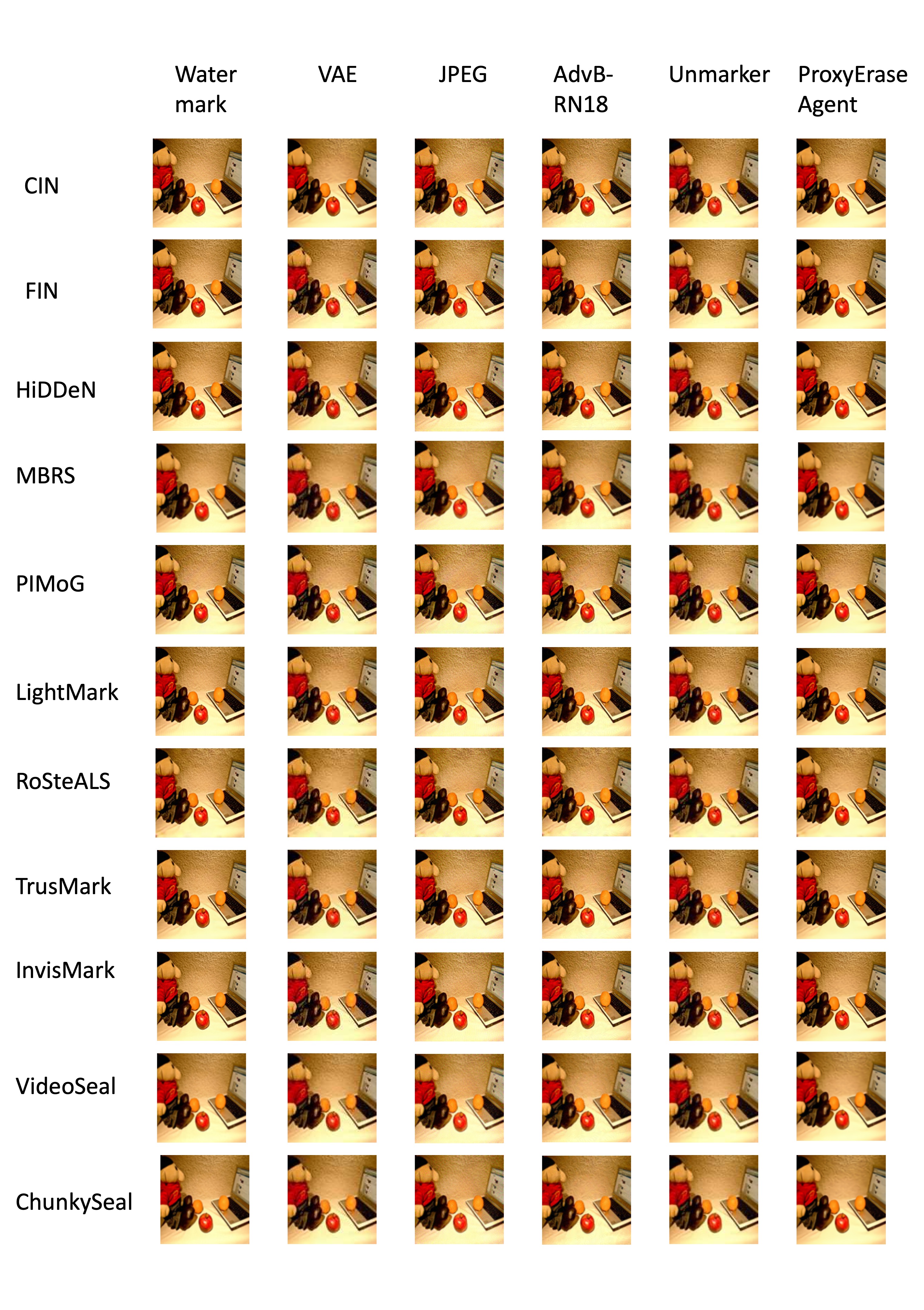}
    \caption{Examples of watermark image attacks, including JPEG compression, VAE-based reconstruction, AdvB-RN18 adversarial attack, Unmarker adversarial attack, and ProxyEraseAgent. 
    }
    \label{fig:baseline_picture1}
\end{figure}

\end{document}